\documentclass[a4paper,11pt]{article}
\usepackage{jheppub}

\usepackage{graphicx,epsfig}
\usepackage{simplewick}
\usepackage{slashed}
\usepackage{color}
\usepackage{setspace}
\usepackage{hyperref}

\newcommand{\be}{\begin{equation}}
\newcommand{\ee}{\end{equation}}
\newcommand{\MS}{\overline{\mathrm{MS}}}
\newcommand{\MC}{\mathrm{MC}}

\newcommand{\latt}{\mathrm{latt}}
\newcommand{\nn}{\nonumber}
\newcommand{\al}{\alpha}
\newcommand{\order}{\mathcal{O}}

\newcommand{\eq}[1]{Eq.~\eqref{#1}}

   \graphicspath{{./figs/}}
\usepackage{color}

\usepackage{relsize}
\newcommand{\babar}{{\mbox{\slshape B\kern-0.1em{\smaller A}\kern-0.1em
            B\kern-0.1em{\smaller A\kern-0.2em R}}}
\def\MSbar{\relax\ifmmode\overline                        
            {\rm MS}\else{$\overline{\rm MS}${ }}\fi}     
           }                                              
\def\MSbar{\relax\ifmmode\overline                        
            {\rm MS}\else{$\overline{\rm MS}${ }}\fi}     
\def\1{\hbox{{1}\kern-.25em\hbox{l}}}
\def\be{\begin{equation}}
\def\ee{\end{equation}}
\def\bea{\begin{eqnarray}}
\def\eea{\end{eqnarray}}
\def\bear{\begin{array}}
\def\eear{\end{array}}

\def\als{\alpha_{s}}
\def\al{\alpha}
\def\nn{\nonumber}

\begin{document}
\title{Hyperasymptotic approximation to the plaquette and determination of the gluon condensate}
\author[a]{Cesar Ayala,}
\author[b]{Xabier Lobregat,}
 \author[b,c]{and Antonio Pineda}
\affiliation[a]{Department of Physics, Universidad T{\'e}cnica Federico
Santa Mar{\'\i}a (UTFSM),  
Casilla 110-V, 
Valpara{\'\i}so, Chile}
\affiliation[b]{Institut de F\'\i sica d'Altes Energies (IFAE), The Barcelona Institute of Science and Technology, Campus UAB, 08193 Bellaterra (Barcelona), Spain}
\affiliation[c]{Grup de F\'{\i}sica Te\`orica, Dept. F\'\i sica, Universitat Aut\`onoma de Barcelona,
E-08193 Bellaterra, Barcelona, Spain}


\abstract{
We give the hyperasymptotic expansion of the plaquette with a precision that includes the terminant associated to the leading renormalon. Subleading effects are also considered. The perturbative series is regulated using the principal value prescription for its Borel integral. We use this analysis to give a determination of the gluon condensate in SU(3) pure gluodynamics that is independent of the scale and renormalization scheme used for the coupling constant: $\langle G^2 \rangle_{\rm PV} (n_f=0)=3.15(18)\; r_0^{-4}$.
}
\maketitle

\section{Introduction}

The expectation value
of the plaquette calculated in Monte Carlo
(MC) simulations in lattice regularization with the
standard Wilson gauge action~\cite{Wilson:1974sk} reads
\begin{equation}
\langle P\rangle_{\mathrm{MC}}=\frac{1}{N^4}\sum_{x\in\Lambda_E}\langle P_x\rangle\,,
\end{equation}
where $\Lambda_E$ is a Euclidean spacetime lattice and
\begin{equation}
P_{x,\mu\nu}=1-\frac{1}{6}\mathrm{Tr}\left(U_{x,\mu\nu}+U_{x,\mu\nu}^{\dagger}\right)\,.
\end{equation}
For details on
the notation see Ref.~\cite{Bali:2014fea}. This quantity can be computed using the operator product expansion 
\be
\label{OPE}
\langle P \rangle_{\mathrm{MC}}=
\sum_{n=0}^{\infty}p_n\al^{n+1}
+\frac{\pi^2}{36}C_{\rm G}(\al)\,a^4\langle G^2 \rangle+\mathcal{O}\left(a^6\right)\,,
\ee
where $a$ denotes the lattice spacing, and 
\begin{align}
\label{eq:GC}
\langle G^2\rangle\equiv -\frac{2}{\beta_0}\left\langle\Omega\left| \frac{\beta(\alpha)}{\alpha}
G_{\mu\nu}^aG_{\mu\nu}^a\right|\Omega\right\rangle
=
\left\langle\Omega\left|  \left[1+\mathcal{O}(\alpha)\right]\frac{\alpha}{\pi}
G_{\mu\nu}^aG_{\mu\nu}^a\right|\Omega\right\rangle
\end{align}
 is
the so-called non-perturbative gluon condensate~\cite{Vainshtein:1978wd}, which, under some conditions, it is expected to scale like $\Lambda^4$ where (in an arbitrary scheme)
\be
\label{eq:betafun}
\Lambda=\mu\exp\left\{-\left[\frac{2\pi}{\beta_0\alpha(\mu)}
+b
\ln\left(\frac12 \frac{\beta_0\alpha(\mu)}{2\pi}\right)
+\sum_{j\geq 1}
s_j\,(-b)^j\!\left(\frac{\beta_0\alpha(\mu)}{2\pi}\right)^{\!j}\right]\right\}
\ee
with
\begin{equation}
b=\frac{\beta_1}{2\beta_0^2}\,,\quad
s_1=\frac{\beta_1^2-\beta_0\beta_2}{4b\beta_0^4}\,,\quad
s_2=\frac{\beta_1^3-2\beta_0\beta_1\beta_2+\beta_0^2\beta_3}{16b^2\beta_0^6}\,,
\end{equation}
and so on.

The Wilson coefficient multiplying the gluon condensate is proportional to the inverse of the beta function of $\alpha$~\cite{DiGiacomo:1990gy,DiGiacomo:1989id}. 
This fixes the Wilson coefficient exactly:
\begin{align}
C_{\rm G}(\al)&=1+\sum_{k\geq 0}c_k\al^{k+1}=
\label{CP}
-\frac{\beta_0\al^2}{2\pi\beta(\al)}
\\
\nn
&=1-\frac{\beta_1}{\beta_0}\frac{\al}{4\pi}
+\frac{\beta_1^2-\beta_0\beta_2}{\beta_0^2}\left(\frac{\al}{4\pi}\right)^2
-\frac{\beta_1^3-2\beta_0\beta_1\beta_2+\beta_0^2\beta_3}{\beta_0^3}\left(\frac{\al}{4\pi}\right)^3
+\order(\al^4)
\,.
\end{align}
Note that $C_{\rm G}(\al)$ is
scheme-dependent not only through $\alpha$, but also
explicitly, due to its dependence on the higher $\beta$-function
coefficients: $\beta_2$, etc..
The $c_k$ 
depend on the $\beta_i$ with $i\leq k+1$ via \eq{CP}.
For $j\leq 3$ the coefficients $\beta_j$  are known in the Wilson action lattice
scheme. $\beta_2^{\latt}$ has been
computed diagrammatically~\cite{Luscher:1995np,Christou:1998ws,Bode:2001uz}.
The value for $\beta_3^{\latt}$ that we use~\cite{Bali:2013qla} is an update of \cite{Bali:2013pla}, 
and was obtained by calculating the
normalization of the leading renormalon of the pole mass, and
then assuming the corresponding $\MS$-scheme 
expansion to follow its asymptotic behaviour
from orders $\als^4$ onwards. 
Similar estimates, $\beta_3^{\latt}\approx
-1.37\times 10^6$ up to $\beta_3^{\latt}\approx
-1.55\times 10^6$, were found in Ref.~\cite{Guagnelli:2002ia}
using a very different method. For convenience, we also write the expansion coefficients $c_k$
defined in \eq{CP} in terms of the constants that appear in \eq{eq:betafun}:
\begin{equation}
\label{eq:relatecs}
c_0=-b\frac{\beta_0}{2\pi}\,,\quad
c_1=s_1b\left(\frac{\beta_0}{2\pi}\right)^{\!2}\,,\quad
c_2=-2s_2b^2\left(\frac{\beta_0}{2\pi}\right)^{\!3}\,.
\end{equation}

The perturbative sum and the leading nonperturbative correction in \eq{OPE} are ill-defined. The reason is that the perturbative series is divergent due
to renormalons~\cite{Hooft} (for a review see \cite{Beneke:1998ui}) and other, subleading, instabilities.
This makes any determination of $\langle G^2\rangle$ ambiguous,
unless we define how to truncate or how to
approximate the perturbative series. Any reasonable definition consistent with
$\langle G^2\rangle \sim \Lambda^4$ can only be given if the asymptotic
behaviour of the perturbative series is under control.
This has only been achieved recently~\cite{Bali:2014fea},
where the perturbative expansion of the plaquette was
computed up to $\mathcal{O}(\al^{35})$. The observed
asymptotic behaviour was in full compliance with renormalon
expectations, with successive contributions starting
to diverge for orders around $\al^{27}$--$\al^{30}$ within
the range of couplings $\al$ typically employed in present-day
lattice simulations. 

Extracting the gluon condensate from the
average plaquette was
pioneered in Refs.~\cite{Di Giacomo:1981wt,Kripfganz:1981ri,DiGiacomo:1981dp,Ilgenfritz:1982yx}, and
many attempts followed during the next decades,
see, e.g., Refs.~\cite{Alles:1993dn,DiRenzo:1994sy,Ji:1995fe,DiRenzo:1995qc,Burgio:1997hc,Horsley:2001uy,Rakow:2005yn,Meurice:2006cr,Lee:2010hd,Horsley:2012ra}. Nevertheless, 
they suffered from insufficiently high perturbative orders and,
in some cases, also finite volume
effects. The failure to make a controlled contact to the asymptotic regime 
prevented a reliable lattice determination of $\langle G^2\rangle$, where one could quantitatively assess the error associated to these determinations. This problem was first solved in \cite{Bali:2014sja}. In such paper, for the first time, the perturbative sum was computed with superasymptotic accuracy for the case of 4 dimensional SU(3) gluodynamics. This allowed to obtain a reliable determination of $\langle G^2 \rangle$ that scaled as $\Lambda^4$. One issue raised was to determine to which extent such a result was independent of the scheme used for the coupling constant. The answer to this question can be given within the general framework of hyperasymptotic expansions of renormalizable quantum field theories, as developed in \cite{Ayala:2019uaw,Ayala:2019hkn,Ayala:2019lak,Ayala:2020odx} (see \cite{BerryandHowls,Boyd99} for the original works in the context of ordinary differential equations). This answer was indeed given in \cite{Ayala:2019uaw}, where it was concluded that the error of using the superasymptotic approximation to the perturbative sum was of ${\cal O}(\sqrt{\alpha(1/a)} Z_P\Lambda^4)$, where $Z_P$ is the normalization of the leading renormalon. This error then sets the parametric precision of the determination of the gluon condensate using the superasymptotic approximation. Note that the scheme dependence of $Z_P$ and $\Lambda^4$ cancels each other. Therefore, the only remaining/leading scheme/scale dependence of the error is due to the $\sqrt{\alpha(1/a)}$ prefactor. It is the purpose of this paper to revisit such analysis and to encode such a result in the more general framework of the hyperasymptotic expansions. In particular, we will reach a better (or more robust) accuracy by incorporating the leading terminant. We will also discuss subleading effects. We confirm that the result we obtain is independent of the scheme/scale used for the renormalization of coupling constant (up to terms that are higher order than the accuracy reached by the hyperasymptotic approximation). 

In order to carry out the above program, the first step is to regularize the perturbative sum, which we do using the Principal Value (PV) prescription. Only after regularizating the perturbative sum, the definition of the gluon condensate is unambiguous and the operator product expansion of the plaquette reads
\be
\label{OPEPV}
\langle P \rangle_{\mathrm{MC}}=
S_{\rm PV}
+\frac{\pi^2}{36}C_{\rm G}(\al)\,a^4\langle G^2  \rangle_{\rm PV}+\mathcal{O}\left((a\Lambda)^6\right)\,.
\ee
This expression is, in practice, formal, as the exact expression of $S_{\rm PV}$ is not known. This would require the exact knowledge of the Borel transform of the perturbative sum. Nevertheless, it is possible to obtain an approximate expression of it with a known parametric control of the error using its hyperasymptotic expansion. The accuracy of this expansion is limited from the information we get from perturbation theory. For the case at hand, we have  
\be
\label{eq:SPV}
S_{\rm PV}=
\sum_{n=0}^{N_P}p_n\alpha^{n+1}+\Omega_{G^2}+
\sum_{n=N_P+1}^{N'}[p_n-p_n^{\rm (as)}]\alpha^{n+1}
+\cdots\,,
\ee
where $N'$ is the maximal order in perturbation theory that is included in the perturbative expansion. Within the hyperasymptotic counting, approximating $S_{\rm PV}$ by $S_{\rm P}\equiv 
\sum_{n=0}^{N_P}p_n\alpha^{n+1}$, the first term in \eq{eq:SPV}, corresponds to the superasymptotic approximation, which we label as $(0,N_P)$. Adding $\Omega_{G^2}$ to the superasymptotic approximation corresponds to (4,0) precision in the hyperasymptotic approximation and adding the last term corresponds to $(4,N')$ precision\footnote{The labeling (D,N) in general is defined in Refs. \cite{Ayala:2019hkn,Ayala:2019lak}.}. 

In \eq{eq:SPV}, we take
\be
\label{eq:NP}
N_P=4\frac{2\pi}{\beta_0\al(1/a)}\left(1-c\al(1/a)\right)
\,,
\ee
as the order at which we truncate the perturbative expansion to reach the superasymptotic approximation. By default, we will take the smallest positive value of $c$ that yields an integer value for $N_P$, but we also explore the dependence of the result on $c$. Note that the value of $N_P$ that we use here is slightly different from the value used in \cite{Bali:2014sja} to truncate the perturbative expansion with superasymptotic accuracy. In that reference, such number was named $n_0$ and was determined numerically. We will ellaborate on this difference later. 

The asymptotic expression of the coefficients of the perturbative expansion was worked out in ~\cite{Bali:2014fea}. We repeat it here for convenience
\begin{align}
\label{pn}
p_n^{\rm (as)} &= Z_{P}\,
\left(
\frac{\beta_0}{2\pi d}\right)^{\!n}
\frac{\Gamma(n+1+db)}{\Gamma(1+db)}
\left\{
1+\frac{db}{n+db}\,b_1
\right.
\\\nn
&\qquad
\left.
+\frac{(db)^2}{(n+db)(n+db-1)}\,
b_2
+
{\cal O}\left(\frac{1}{n^3}\right)
\right\}
\,.
\end{align}
Note that the parameters $b_1$ and $b_2$:
\begin{align}
\label{eq:relatec0}
b_1&=ds_1+\frac{2\pi c_0}{\beta_0b}=ds_1-1\,,\\
\label{eq:relatec1}
b_2&=
\frac{4\pi^2c_1}{\beta_0^2b^2}
+ds_1\left(\frac{ds_1}{2}+\frac{2\pi c_0}{\beta_0b}\right)
-ds_2=ds_1\left(\frac{ds_1}{2}-1+\frac{1}{db}\right)-ds_2\,,
\end{align}
that describe the leading
pre-asymptotic corrections depend on
the expansion coefficients $c_0$ and $c_1$,
defined in \eq{CP},
of the Wilson coefficient
of the gluon condensate.

$\Omega_{G^2}$ is the terminant associated to the leading renormalon of the plaquette. It can be easily obtained from the general formulas given in \cite{Ayala:2019uaw}. It reads
\be
	\label{OmegaExact}
	\Omega_{G^2}=\Delta \Omega(4b)+b_1\Delta \Omega(4b-1)+w_2\Delta \Omega(4b-2)+\cdots 
\,,
\ee
where we take $\Delta \Omega(db)$ from Eq. (33) of \cite{Ayala:2019uaw} taking $\gamma=0$, and
$$
w_2=\frac{4 b_2 b}{4b-1}
\,.
$$
$\Omega_{G^2}$ can also be written in the following way
\be
	\label{OmegaLambda}
	\Omega_{G^2}=
	\sqrt{\al(1/a)}K^{(P)}a^4\Lambda^4
	\bigg(
		1
		+K_{1}^{(P)}\al(1/a)
		+K_{2}^{(P)}\al^2(1/a)
		+\mathcal{O}(\al^3(1/a))\bigg)
\,,
\ee
or
\begin{align}
	\label{OmegaExp}
	\Omega_{G^2}=&
	\sqrt{\al(1/a)}K^{(P)}
	e^{-\frac{8\pi}{\beta_0 \al(1/a)}}
	\left(\frac{\beta_0\al(1/a)}{4\pi}\right)^{-4b}
	\bigg(
		1
		+\bar K_{1}^{(P)}\al(1/a)
		\nonumber
		\\
		&+\bar K_{2}^{(P)}\al^2(1/a)
		+\mathcal{O}(\alpha^3(1/a))\bigg)
\,,
\end{align}
where 
\begin{align}
	&K^{(P)}=
	\frac{-Z_P}{\Gamma(1+4b)}
	2^{2+4b}\pi\beta_0^{-1/2}
	\left(-\eta_c+\frac{1}{3}\right)
\,,
	\\
	&\bar K_{1}^{(P)}=
	\frac{\beta_0/(4\pi)}{-\eta_c+\frac{1}{3}}
	\bigg[
		-4bb_1\left(\frac{1}{2}\eta_c+\frac{1}{3}\right)
		-\frac{1}{12}\eta_c^3
		+\frac{1}{24}\eta_c
		-\frac{1}{1080}\bigg]
\,,	\\
	&K_{1}^{(P)}=\bar K_{1}^{(P)}-\frac{2b \beta_0 s_1}{\pi}
\,,	\\
	&\bar K_{2}^{(P)}=
	\frac{\beta_0^2/(4\pi)^2}{-\eta_c+\frac{1}{3}}
	\bigg[
		-4 w_2(4b -1)b\left(\frac{1}{4}\eta_c+\frac{5}{12}\right)
		\nonumber
		\\
		&\qquad
		+4b_1 b\left(
			-\frac{1}{24}\eta_c^3
			-\frac{1}{8}\eta_c^2
			-\frac{5}{48}\eta_c
			-\frac{23}{1080}\right)
		-\frac{1}{160}\eta_c^5
		-\frac{1}{96}\eta_c^4
		+\frac{1}{144}\eta_c^3
		\nonumber
		\\
		&\qquad
		+\frac{1}{96}\eta_c^2
		-\frac{1}{640}\eta_c
		-\frac{25}{24192}\bigg]
\,,	\\
	&K_{2}^{(P)}=
	\frac{1}{8\pi^2}
	\big(
		8\pi^2\bar K_{2}^{(P)}
		-16b\pi s_1\beta_0\bar K_{1}^{(P)}
		+16b^2s_1^2\beta_0^2
		+8b^2s_2\beta_0^2\big)
\,,
\end{align}
where $\eta_c\equiv-4b+\frac{8\pi}{\beta_0}c-1$.

The value of $Z_P$ was determined approximately (for $n_f=0$) in \cite{Bali:2014fea}:
\be
Z_P=(42\pm 17)\times 10^4
\,.
\ee
This is the value we will use in this paper. Actually, its error will give the major source of uncertainty in the determination of $\Omega_{G^2}$, of the order of 40\%. The other source of error is due to the fact that only approximate expressions are  available for $\Omega_{G^2}$ (see \eq{OmegaExact},  \eq{OmegaLambda}, and  \eq{OmegaExp}), as we do not know the complete set of coefficients of the beta function in the lattice scheme. Nevertheless, we can study the convergence pattern of the weak-coupling expansion. We show the results in Table \ref{Tab:Omega} for a representative set of values of $\alpha$ in the interval that we will use later. The first observation is that we observe a very good convergent pattern of the 
weak-coupling expansion of the terminant using \eq{OmegaExp} or \eq{OmegaExact}, consecutive terms quickly become smaller. The latter is obtained using the exact numerical determinations of $\Delta \Omega$. The second observation is that 
the strict weak-coupling expansion used in \eq{OmegaExp} saturates perfectly the exact numerical determination of \eq{OmegaExact} for analogous precision. On the other hand, if we want to use \eq{OmegaLambda} the convergence is not 
good. We have to go to $\beta$-values ($\beta\equiv 3/(2\pi\alpha)$) rather larger than 6 to get decent accuracy. What lies behind is the fact that $\Lambda_{\rm latt}$ is not well approximated by its weak coupling expansion at low orders (see the 
discussion in \cite{Bali:2014sja} and \cite{Ayala:2019hkn}). This bad convergence of the weak coupling expansion of $\Lambda_{\rm latt}$ has to be compensated by a bad convergence of the ${\cal O}(\alpha)$ corrections of the weak coupling expansion in \eq{OmegaLambda}. A similar behavior, but less severe, was observed in \cite{Ayala:2019hkn}. In 
that paper, the analogous to the Wilson coefficient $C_G$ was 1. Nevertheless, the main point is that now the power of $\Lambda$ is four, whereas in \cite{Ayala:2019hkn}, the power of $\Lambda$ was one. This makes that the relatively bad convergent behavior observed in \cite{Ayala:2019hkn} gets amplified by a factor of four here.\footnote{It has a theoretical interest to study the behavior of $\Omega_{G^2}$ in the $\MS$ scheme. In this scheme, $\Lambda_{\rm MS}$ is well approximated by its weak-coupling expansion. Consequently, in this scheme, \eq{OmegaExact}, \eq{OmegaExp}, and \eq{OmegaLambda} converge well. It is also interesting to study these equations with $n_f \not=0$, in view of future full QCD analyses, which will involve the incorporation of active massless fermions (for preliminary studies see \cite{DelDebbio:2018ftu}). As expected, one observes a better convergence of the weak-coupling expansion taking $n_f=3$ than $n_f=0$.}
 Therefore, in the following, we will always use \eq{OmegaExp} as our approximated expression for $\Omega_{G^2}$, as it produces a nicely convergent series with a controlled scheme dependence, as the weak coupling expansion is organized in terms of a single parameter: $\alpha$. The error associated to truncating the expansion in \eq{OmegaExp} is estimated by observing the convergent pattern of the LO, NLO and NNLO results in Table \ref{Tab:Omega}. From LO to NLO, in the worst cases, the differences are close but below 50\%, and from NLO to NNLO the differences are below 10\%. One could then expect the NNNLO contribution to be at the level of few percent, which can be neglected all together in comparison with the $\sim 40\%$ error associated to $Z_P$.
 
\medskip
\begin{table}
\centering
\label{Tab:Omega}
\caption{A representative set of values of $\Omega_{G^2}$ using \eq{OmegaExact} (exact) and \eq{OmegaExp} (exp) 
truncated at different order in their respective expansions.}
\resizebox{\textwidth}{!}{
\begin{tabular}{|c|c|c|c|c|c|c|c|}
\hline
$\beta$  & $N_P$ & $\Omega _{\text{LO}}^{(\text{exp})}\!\!\times\!\! 10^5$ & $\Omega _{\text{LO}}^{(\text{exact})}\!\!\times\!\! 10^5$ & $\Omega _{\text{NLO}}^{(\text{exp})}\!\!\times\!\! 10^5$ & $\Omega _{\text{NLO}}^{(\text{exact})}\!\!\times\!\! 10^5$ & $\Omega _{\text{NNLO}}^{(\text{exp})}\!\!\times\!\! 10^5$ & $\Omega _{\text{NNLO}}^{(\text{exact})}\!\!\times\!\! 10^5$ \\  \hline
 5.8 & 27 & -16.23 & -16.50 & -23.06 & -23.07 & -24.31 & -24.30 \\  \hline
 6. & 28 & -6.611 & -6.723 & -9.344 & -9.349 & -9.888 & -9.886 \\  \hline
 6.2 & 29 & -2.689 & -2.735 & -3.781 & -3.783 & -4.013 & -4.011 \\  \hline
 6.4 & 30 & -1.092 & -1.111 & -1.528 & -1.529 & -1.625 & -1.624 \\  \hline
 6.6 & 31 & -0.4426 & -0.4505 & -0.6165 & -0.6170 & -0.6571 & -0.6561 \\  \hline
\end{tabular}}
\end{table}

\section{Determination of the gluon condensate}
\label{Sec:Gluon}

Following the notation of \cite{Bali:2014sja}, we determine the gluon condensate from the following equation: 
\be
\label{eq:Fit}
\langle G^2 \rangle_{\rm PV} = \frac{36 C_G^{-1}}{\pi^2 a^4}
\left[ \langle P \rangle_{\rm MC} -S_{\rm PV}
\right]
\,.
\ee
If $S_{\rm PV}$ and $\langle P \rangle_{\rm MC}$ were known exactly, this equality is expected to hold up to corrections of ${\cal O}(a^2\Lambda^2)$. Nevertheless, neither $S_{\rm PV}$ nor $\langle P \rangle_{\rm MC}$ are known exactly. On top of that, we have to account for the fact that $C_G^{-1}$ and the relation between $a$ and $\beta$ are also known in an approximated way. We now discuss how we determine them and their associated individual errors. 

\begin{center}
\begin{figure}
\includegraphics[width=0.95\textwidth]{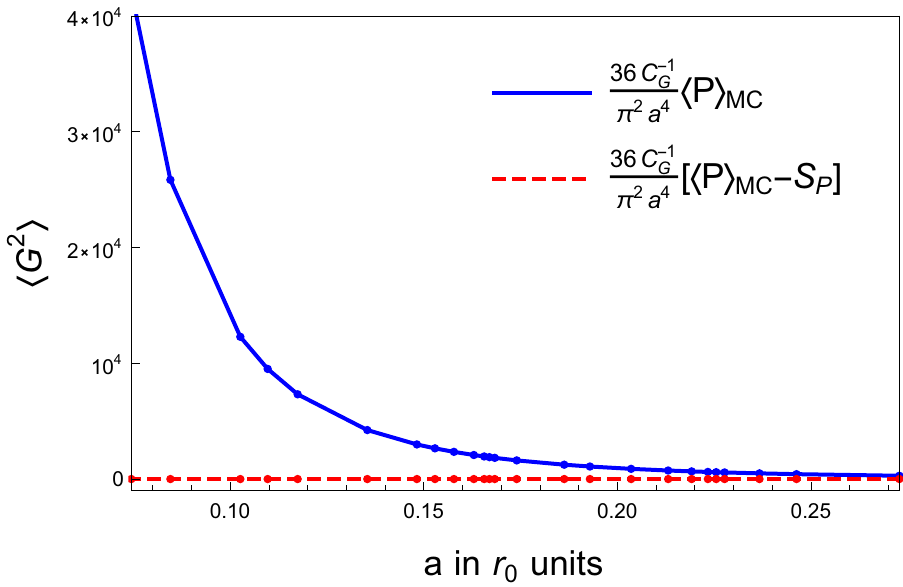}
\caption{$ \frac{36 C_G^{-1}}{\pi^2 a^4} \langle P \rangle_{\rm MC}$ (continuous blue line) and $ \frac{36 C_G^{-1}}{\pi^2 a^4}\left[ \langle P \rangle_{\rm MC} -S_{P} \right]$ (dashed red line). The second line is basically indistinguishable with respect to zero with the scale resolution of this plot. The statistical errors are smaller than the size of the points.}
\label{Fig:cancellation}
\end{figure}
\end{center}

We take the MC data from \cite{Boyd:1996bx}. Similarly to what was done in \cite{Bali:2014sja}, in this analysis
we restrict ourselves to the more precise
$N=32$ data and, to keep finite volume effects under control,
to $\beta\leq 6.65$. We also 
limit ourselves to $\beta\geq 5.8$ to avoid large
$\mathcal{O}(a^2)$ corrections. 
At very large $\beta$-values, obtaining meaningful results becomes challenging
numerically: the individual errors both of $\langle P\rangle_{\MC}(\al)$
and of $S_{\rm PV}(\al)$ somewhat decrease with increasing $\beta$.
However, there is a very strong cancellation between these two terms, in
particular at large $\beta$-values, since this difference
decreases with $a^{-4}
\sim\Lambda_{\latt}^4\exp(16\pi^2\beta/33)$ on dimensional
grounds, while $\langle P\rangle_{\MC}$ depends only logarithmically
on $a$. We illustrate this cancellation in Fig. \ref{Fig:cancellation}. 

Equation~(\ref{eq:betafun}) is not accurate enough in the lattice scheme
for the $\beta$-values used in this paper.
Instead, we employ the
phenomenological parametrization of Ref.~\cite{Necco:2001xg}
($x=\beta-6$)
\begin{align}
\label{eq:Necco}
a=r_0\exp\left(-1.6804-1.7331x+0.7849x^2-0.4428x^3\right)\,,
\end{align}
obtained by
interpolating non-perturbative lattice simulation results.\footnote{For a more detailed comparison of this phenomenological parameterization and its weak coupling approximation see the discussion in \cite{Bali:2014sja} and \cite{Ayala:2019hkn}.}
Equation~(\ref{eq:Necco}) was reported to be valid within an accuracy varying
from 0.5\% up to 1\% in the
range~\cite{Necco:2001xg} $5.7\leq\beta\leq 6.92$, which includes the range $\beta \in [5.8,6.65]$ we use in this paper. 
This range corresponds to
$(a/r_0)^4 \in [3.1\times 10^{-5},5.5\times 10^{-3}]$, and covers more than
two orders of magnitude, i.e. in units of energy, we use lattice data in the region $1/a \sim (3.66 \; r_0^{-1} \div 13. 42 \;  r_0^{-1})$. 

For the inverse Wilson coefficient
\begin{align}
C^{-1}_{\rm G}(\al)&=
\label{CPinverse}
-\frac{2\pi\beta(\al)}{\beta_0\al^2}
\\
\nn
&=1+\frac{\beta_1}{\beta_0}\frac{\al}{4\pi}
+\frac{\beta_2}{\beta_0}\left(\frac{\al}{4\pi}\right)^2
+\frac{\beta_3}{\beta_0}\left(\frac{\al}{4\pi}\right)^3
+{\cal O}(\al^4)
\,,
\end{align}
the corrections to
$C_{\rm G}=1$ are small. However, the $\mathcal{O}(\alpha^2)$ and
$\mathcal{O}(\alpha^3)$ terms are of similar sizes. We will
account for this uncertainty in our error budget.

\medskip

\begin{figure}
\centering
\includegraphics[width=0.95\textwidth]{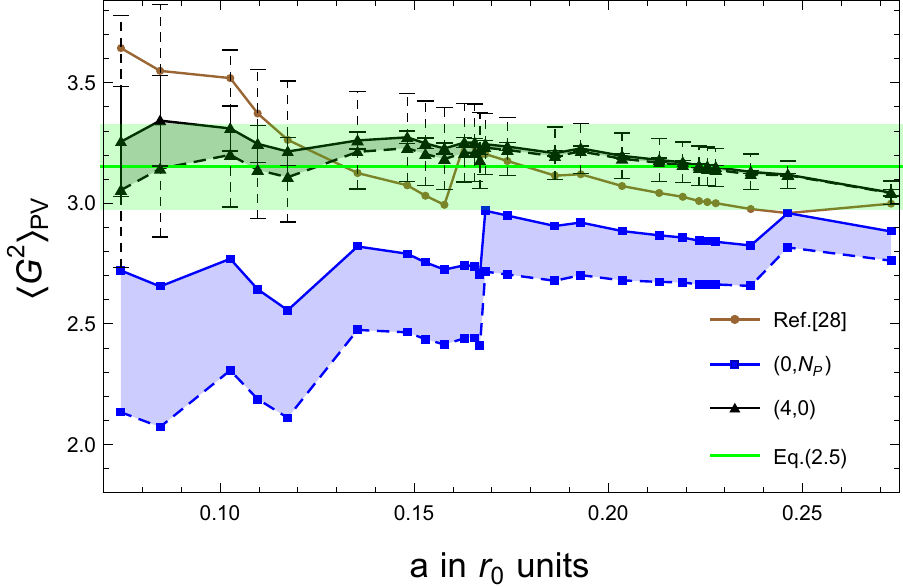}
\caption{Gluon condensate with superasymptotic approximation $(0,N_P)$ and with hyperasymptotic accuracy $(4,0)$. In both cases, for each corresponding $\beta$, we show the value obtained for the gluon condensate with the values of $N_P$ using the smallest positive (upper line) and negative (lower line) value of $c$ that yields an integer value of $N_P$. For the hyperasymptotic approximation with $c$ positive we also show the statistical errors of the MC determination of the plaquette (inner error) and its combination in quadrature with the statistical error of the partial sum (outer error). We also show the superasymptotic approximation obtained in \cite{Bali:2014sja} truncating at the minimal term determined numerically. The horizontal green band and its central value are our final prediction, and the associated error, for the gluon condensate displayed in \eq{G240}.}
\label{Fig:G2}
\end{figure}

We now turn to $S_{\rm PV}(\al)$. As we have mentioned above, we compute it using the hyperasymptotic expansion. This introduces a parametric error according to the order we truncate this expansion. On top of that, the coefficients $p_n$, obtained in Ref.~\cite{Bali:2014fea}, are not known exactly. They carry statistical errors,
and successive orders are correlated. Using the covariance matrix, also
obtained in Ref.~\cite{Bali:2014fea}, the statistical error of $S_P(\al)$ can
be calculated. In that reference, coefficients $p_n(N)$ were first
computed on finite volumes of $N^4$ sites and subsequently extrapolated
to their infinite volume limits $p_n$. This extrapolation
is subject to parametric uncertainties that need to be
estimated. We follow Ref.~\cite{Bali:2014fea}
and add the differences between determinations
using $N\geq \nu$ points for $\nu=9$ (the central values)
and $\nu=7$ as systematic errors to our statistical errors. This is the same error analysis as the one used in \cite{Bali:2014sja}. We emphasize though, that the order we truncate the perturbative series, $N_P$, is different from the one used in \cite{Bali:2014sja} (which was named $n_0$ in this reference). The difference between both determinations gives an estimate of the parametric error of the determination of $S_{\rm PV}(\al)$ by using the superasymptotic approximation $S_P$. The magnitude of $\Omega_{G^2}$ gives an alternative estimate of the error associated to the truncation of the hyperasymptotic approximation. It is also interesting to see the magnitude of changing $N_P$ by one unit by fine tunning $c$ from the smallest positive value that yields an integer value of $N_P$ to the smallest (in modulus) negative value that yields an integer value of $N_P$. Typically this yields slightly smaller errors. We illustrate this discussion in Fig. \ref{Fig:G2}. All these error estimates scale with the parametric uncertainty predicted by theory $\sim {\cal O}(e^{-4\frac{2\pi}{\beta_0 \al(1/a)}}) \sim {\cal O}(a^4\Lambda^4)$ times $\sqrt{\alpha}$ (see the discussion in \cite{Ayala:2019hkn,Ayala:2019lak}).

If we increase the accuracy of the hyperasymptotic expansion by adding the terminant $\Omega_{G^2}$ to the superasymptotic approximation, the parametric error decreases, and the accuracy reached is (4,0) (note that the statistical error does not change). With this accuracy, the parametric error is $\sim {\cal O}(e^{-4\frac{2\pi}{\beta_0 \al(1/a)}(1+\log(3/2))}) \sim  {\cal O}((a\Lambda)^{4(1+\log(3/2))})$ (see the discussion in \cite{Ayala:2019hkn,Ayala:2019lak}). 
Note that $4\log(3/2) \simeq 1.6 <2$. Therefore, these effects are parametrically more important than the next nonperturbative power corrections. Compared with the typical size of the terminant $\Omega_{G^2}$, these effects are suppressed by a factor of order $\sim {\cal O}((a\Lambda)^{4\log(3/2))})$. In the energy range we do the fits, this yields suppression factors in the range $((a\Lambda_{\MS})^{4\log(3/2)}) \in (0.007
,
0.05)$, where we have taken $\Lambda=\Lambda_{\MS}$ to be more conservative. This discussion can be affected by powers of $\alpha$. It is expected that there is an extra suppression factor of $\alpha^{3/2}$ (as $\sqrt{\al}$ is already included in the terminants the real suppression factor would be of order $\alpha$). Depending on the scheme, the size of this extra factor is different. In any case, they go in the direction to make the estimate of the error smaller. We will not dwell further in this discussion of the parametric error of the (4,0) hyperasymptotic accuracy, because we only approximately know $\Omega_{G^2}$ and its error will hide the signal of these ${\cal O}((a\Lambda)^{4(1+\log(3/2))})$ effects. For $\Omega_{G^2}$ we use the analytic expression in \eq{OmegaExp} truncated at ${\cal O}(\al^2)$.
The error of this expression comes from $Z_P$, and from the truncation of the weak coupling expansion of the terminant. The largest source of error comes from $Z_P$. Due to its size, this error overwhelms  the parametric error associated to higher-order terms in the hyperasymptotic expansion. 

Irrespective of the discussion of the error of the (4,0) accuracy, it is nice to see that adding the terminant to the superasymptotic expression makes the jumps that we had with the superasymptotic approximation disappear. Adding the terminant also makes the resulting curve flatter. The dependence in $N_P$ (or in other words $c$) gets much milder too. We illustrate all this in Fig. \ref{Fig:G2}.

\medskip

\begin{figure}
\centering
\includegraphics[width=0.67\textwidth]{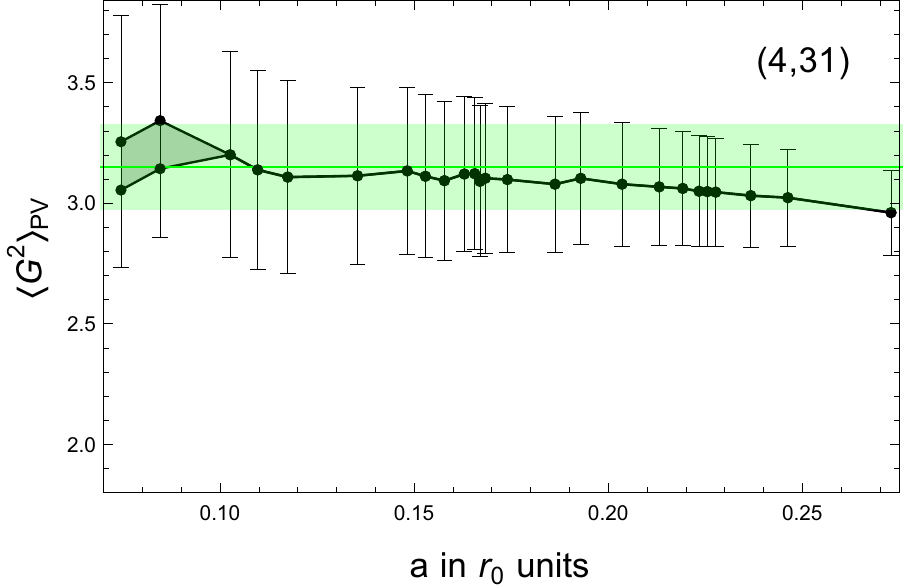}
\includegraphics[width=0.67\textwidth]{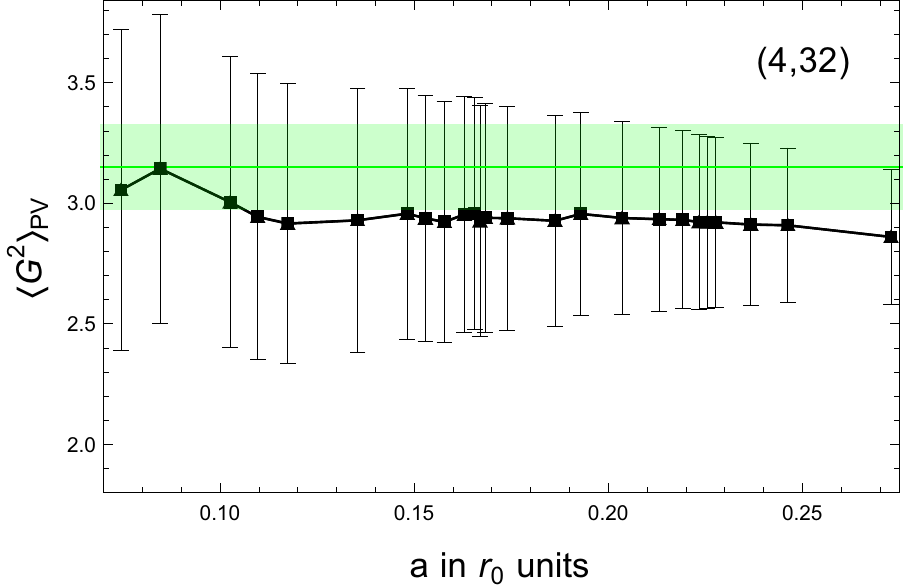}
\includegraphics[width=0.67\textwidth]{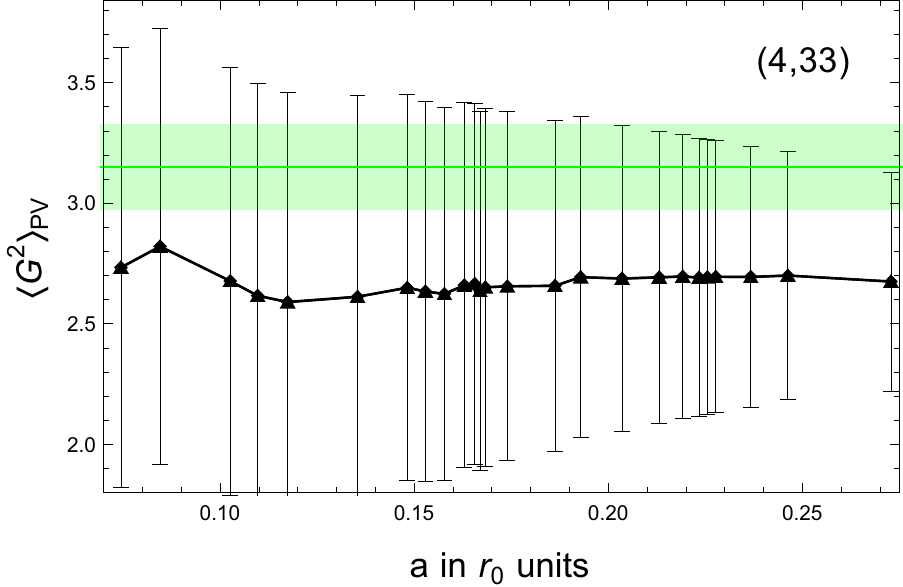}
\caption{Gluon condensate with hyperasymptotic accuracy $(4,N')$ for $N'=31$ (upper pannel), $N'=32$ (middle pannel) and $N'=33$ (lower pannel). In all cases, for each corresponding $\beta$, we show the value obtained for the gluon condensate with the values of $N_P$ using the smallest positive (upper line) and negative (lower line) value of $c$ that yields an integer value of $N_P$. The error is the statistical error of the MC determination of the plaquette  and of the perturbative sum combined in quadrature. The horizontal green band and its central value are our final prediction, and the associated error, for the gluon condensate displayed in \eq{G240}.}
\label{Fig:N'}
\end{figure}

In principle, we know perturbation theory to orders high enough to include the last term written in \eq{eq:SPV} and reach $(4,N')$ accuracy. Nevertheless, we find that the errors of $p_n$ for large $n$ hide the signal. We show in Fig. \ref{Fig:N'} how the statistical errors grow as we increase $N'$. On the other hand it is rewarding to see that the dependence in $c$ basically vanishes. We elaborate more on the error analysis of the $(4,N')$ hyperasymptotic approximation in the next section. 

\subsection{Fit}

We now implement the discussion of the error of the previous section to \eq{eq:Fit} for the different truncations of the hyperasymptotic expansion, and for the associated determination of the gluon condensate. 

The statistical errors of the fits are those of the MC determination of $\langle P \rangle_{\rm MC}$ and the statistical error of $S_{\rm PV}$. The latter is generated by the statistical errors of the coefficients $p_n$. As successive orders are correlated, we use the covariance matrix and, by propagation of the error, compute the statistical error of $S_{\rm PV}$. This is the same method followed in \cite{Bali:2014sja} for the superasymptotic approximation. We then combine the statistical error of $\langle P \rangle_{\rm MC}$ and of $S_{\rm PV}$ in quadrature, which is then used to generate the fits. We show the size of these two different statistical errors in Fig. \ref{Fig:G2}. 

We now turn to systematic uncertainties. One is the infinite volume extrapolation of the coefficients $p_n$ discussed before. Another source of systematic errors is the possible existence of ${\cal O}(a^2 \Lambda^2)$ corrections to the fit. Looking to Fig. \ref{Fig:G2}, within statistical errors, there is no clear  signal of the ${\cal O}(a^2 \Lambda^2)$ in the whole energy range used. Therefore, our default fits will be in the range $\beta \in [5.8,6.65]$ and use the difference with fits in the range $\beta \in [6,6.65]$, as an estimate of these effects. Another source of systematic uncertainties is the incomplete knowledge of $C_G$. We consider the difference between truncating $C_G$ to ${\cal O}(\al^2)$ and to ${\cal O}(\al^3)$ as an estimate of this error. Next, there is a scale error of
about 2.5\%, translating $a$ into units of $r_0$. The other systematic uncertainties are specific to each truncation of the hyperasymptotic approximation used, which we next address. Therefore, 
we then move to discuss the final error of the different orders in the hyperasymptotic approximation. 

\begin{figure}
\centering
\includegraphics[width=0.95\textwidth]{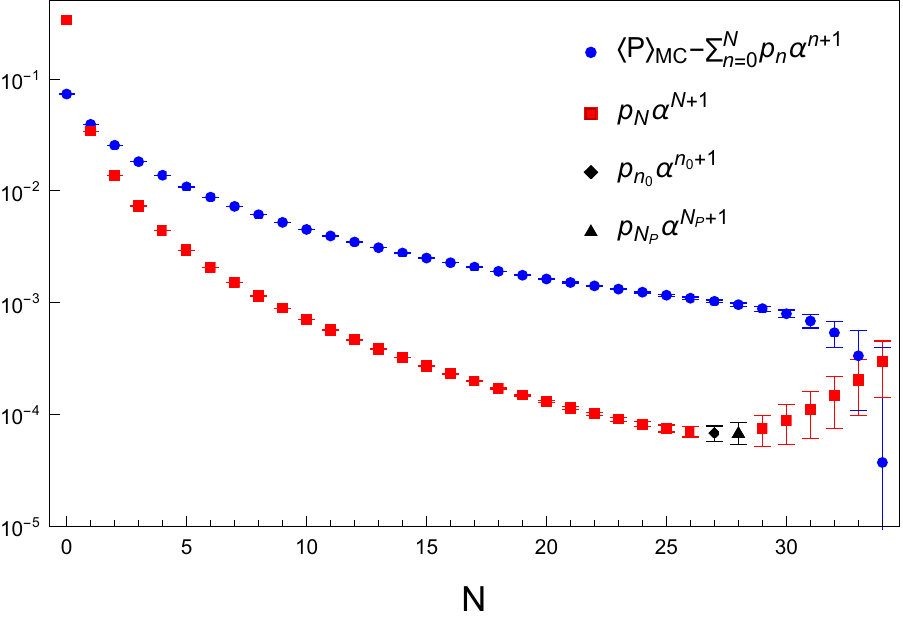}
\caption{We draw $\langle P \rangle_{\mathrm{MC}}-\sum_{n=0}^{N}p_n\al^{n+1}$ (full blue circle points), and $p_N\al^{N+1}$ (full red squares points) for different values of $N$ up to $N=34$ for $\beta=6$. The error of the blue points is the statistical error of the MC simulation and of the sum $\sum_{n=0}^{N}p_n\al^{n+1}$ combined in quadrature (for large $N$ the error of the perturbative sum is dominant). The error displayed here of the perturbative sum does not include the systematic error of the infinite volume extrapolation of the coefficients $p_n$. 
The error displayed for the red points is the complete error (statistical plus systematic combined in quadrature) of the $p_N$ coefficient obtained in \cite{Bali:2014fea} times $\al^{N+1}$. The black diamond stands for the numerical minimal value of $p_N\al^{N+1}$. The black triangle is $p_{N_P}\al^{N_P+1}$ using the smallest positive $c$ that makes $N_P$ to be integer in \eq{eq:NP}. Note that the plus/minus error does not display symmetrically in the plot because of the logarithmic scale.}
\label{Fig:ErrorN}
\end{figure}

{\bf *) Precision $(0,N)$}\\
It is not the purpose of this paper to make a detailed discussion of (how to estimate) the error of the perturbative expansion with precision $(0,N)$ with $N \ll N_P$. Nevertheless, we can not avoid mentioning that there are scenarios where standard ways to estimate the error of the truncation of the perturbative series can underestimate the error. One such standard methods is to take the magnitude of the last term computed (or this quantity multiplied by $\alpha$). Here, working in the lattice scheme, we are indeed in such a situation. The magnitude of the next order of the perturbative series is much smaller than the real error of the computation (the difference between the exact result and the truncated sum to order $N$), as we move away from the first few orders. We illustrate this in Fig. \ref{Fig:ErrorN}. The reason is that each new term of the perturbative series is only marginally smaller than the previous one. This has an important additive effect before reaching the asymptotic regime. Other renormalization schemes of $\alpha$ (closer to the $\MS$ scheme) are expected to work better in this respect (with a smaller ratio between consecutive orders before reaching the asymptotic regime).

{\bf *) Precision $(0,N_P)$}\\
We now want to determine the error of the superasymptotic approximation, which we quantitatively discuss. We first give the number obtained from the fit, as well as the errors:
\be
\label{Eq:G2NP}
\langle G^2 \rangle_{\rm PV} =2.87(2)_{\rm stat.}(6)_{\rm p_n^{\rm ext}}(4)_{\rm range}(8)_{C_G}(7)_{r_0}(28)_{\rm hyp}\; r_0^{-4}=2.87(31) \; r_0^{-4}
\,.
\ee
The first error is the statistical error of the fit. The following errors are systematic. The second error is the error associated to different infinite volume extrapolations of the coefficients $p_n$. Up to this point, the error runs parallel to the error analysis made in \cite{Bali:2014sja}. Nevertheless, unlike in this reference, we do the fit in the range $\beta \in [5.8,6.65]$. If we do the fit in the range $\beta \in [6,6.65]$, as it was done in that reference, the result is -0.04 smaller, a small shift. This is the third error in \eq{Eq:G2NP}. For both ranges the reduced $\chi^2$ are similar: 0.44 and 0.42 for the range $\beta \in [5.8,6.65]$ and the range $\beta \in [6,6.65]$, respectively. On the other hand truncating the partial sum at the numerical minimal term yields, $\langle G^2 \rangle=3.18$ r$_0^{-4}$ with $\chi^2_{\rm red}=0.69$ for the range $\beta \in [6,6.65]$, and $\langle G^2 \rangle=3.05$ r$_0^{-4}$ with  $\chi^2_{\rm red}=1.28$ for the range $\beta \in [5.8,6.65]$. Looking to the points in Fig. \ref{Fig:G2}, we also observe that the remaining $a$ dependence can not be assigned to a specific slope that can be interpreted as an ${\cal O}(a^2\Lambda^2)$ effect, since the sign would flip if taking the sum truncated at the minimal term determined numerically or using \eq{eq:NP} (though the latter seems to yield a flatter curve). Therefore, with the superasymptotic precision, we cannot isolate ${\cal O}(a^2\Lambda^2)$ effects. The fourth error is the difference of the fit truncating $C_G$ to ${\cal O}(\al^2)$ or to ${\cal O}(\al^3)$. The change is significant. This seems to be due to the low convergence of the weak-coupling expansion in the lattice scheme. We have checked that there is convergence (albeit slow) by including higher-order terms of the beta-function using the estimates obtained in \cite{Ayala:2019hkn}. Following \cite{Bali:2014sja}, we assign a 2.5\% error for the conversion from $a$ to $r_0$ units. This is the fifth error in \eq{Eq:G2NP}. 
The last error is the estimate of the higher-order terms in the hyperasymptotic expansion not included in the superasymptotic approximation. This error has been discussed before. This last error is taken as the difference between the fits including or not the leading terminant. This basically gives the same error than considering the difference of doing superasymptotic fits truncating the perturbative sum at the numerical minimal term or using \eq{eq:NP}. Other possible ways to estimate the error (like taking $c$ to be negative such that $N_P$ changes by one unit) give smaller errors. This error is by far the major source of uncertainty of the superasymptotic approximation. In the last equality in \eq{Eq:G2NP} we have combined all these errors in quadrature. 

\medskip

{\bf *) Precision $(4,0)$}\\
We now add the terminant to the superasymptotic approximation. We obtain
\be
\label{G240}
\langle G^2 \rangle_{\rm PV} =3.15(2)_{\rm stat.}(5)_{\rm p_n^{\rm ext}}(9)_{\rm range}(9)_{C_G}(8)_{r_0}(8)_{Z_P}\;r_0^{-4}
=3.15(18) \; r_0^{-4}
\,.
\ee
The error analysis follows to a large extent the error analysis of the superasymptotic approximation. The first error is the statistical error of the fit, with a smaller than one reduced $\chi^2$: 
$\chi^2_{\rm red}=0.43$. 
The following errors are systematic. The second error is the error associated to different infinite volume extrapolations of the coefficients $p_n$. We emphasize again that we do the fits over the whole range $\beta \in [5.8,6.65]$. If we do the fit in the range $\beta \in [6,6.65]$, the result is +0.09 larger with a rather small $\chi^2_{\rm red}=0.019$. This is the third error in \eq{Eq:G2NP}. Having a look to the points in Fig. \ref{Fig:G2}, the remaining $a$ dependence is very small but may point to a small negative slope. If anything, this effect is only visible for the largest distances. At short distances, the $a$ dependence is completely hidden by the errors, which reflects in this very small $\chi^2_{\rm red}$, but even at the largest distances, the errors hide any meaningful signal of these effects. Note that this possible remaining $a$ dependence can be associated to higher-order terms of the hyperasymptotic expansion of $S_{\rm PV}$, which would then scale as ${\cal O}((a\Lambda)^{4(1+\log(3/2))})$ rather than to genuine nonperturbative corrections that would scale as ${\cal O}(a^6\Lambda^6)$. In this respect, it is worth noting that this small slope somewhat tends to disappear as we work with precision $(4,N')$, albeit with a huge error (see Fig. \ref{Fig:N'}). 
The fourth error is the difference of the fit truncating to ${\cal O}(\al^2)$ or to ${\cal O}(\al^3)$ the perturbative expansion of $C_G^{-1}$. The fifth error is the one associated to the conversion from $a$ to $r_0$ units. The last error is the error associated to $Z_P$, the normalization of the leading renormalon. The error of this quantity is heavily correlated to the knowledge of the coefficient $b_2$ (see the discussion in \cite{Bali:2014fea}). Therefore, to estimate this error, we correlate the change of $Z_P$ to setting $b_2=0$. Comparatively to this error, the subleading terms of the weak coupling expansion in \eq{OmegaExp} produce a smaller change and can be neglected. We now discuss the error associated to the truncation of the hyperasymptotic approximation. As discussed before, the leading contributions to the hyperasymptotic expansion of $S_{\rm PV}$ that are not included in the (4,0) precision are expected to scale as ${\cal O}((a\Lambda)^{4(1+\log(3/2))})$ and to be suppressed by a factor  ${\cal O}((a\Lambda)^{4\log(3/2)})$ (times $\alpha$) with respect to the typical size of $\Omega_{G^2}$. This produces corrections at the level of the one/two MeV level. 
Therefore, unlike in the case of the superasymptotic approximation, the errors of the hyperasymptotic approximation (4,0) are small and can be considered to be included in the other errors, in particular in those associated to $Z_P$ and the different ranges we do the fits, as they measure our incomplete knowledge of the perturbative expansion (independently of the truncation of the perturbative expansion in $C_G$). Finally, we combine all the errors in quadrature producing the last equality in \eq{G240}. This is our most precise prediction for $\langle G^2 \rangle_{\rm PV}$, which we display in Fig. \ref{Fig:G2}.  

The central value we obtain does not change much with respect to the central value obtained \cite{Bali:2014sja}. Nevertheless, this is to some extent by accident, as the fit is made over different energy intervals. On the other hand the superasymptotic approximation truncated at the numerical minimal term appears to approach better the central value\footnote{In this respect one could also think of fine tuning the value of $c$ to make $N_P$ to coincide with the numerical minimal term $n_0$.}. We saw that also for the self-energy of the static quark \cite{Ayala:2019hkn}. Nevertheless, the error is larger because the points are more scattered around, and because of the intrinsic inaccuracy of the superasymptotic approximation. In our case, the total error is basically shrunk by a factor 1/2. Note that the statistical error and the error associated to the infinite volume extrapolation of the coefficients are smaller now. The improvement in the quality of the fit can also be observed by the flatter curve we have now.

\begin{figure}
\centering
\includegraphics[width=0.95\textwidth]{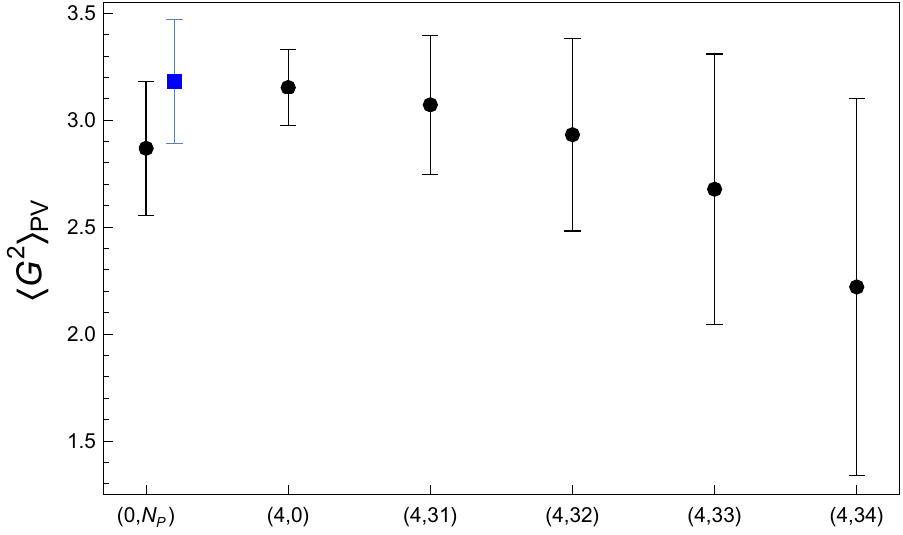}
\caption{Determinations of $\langle G^2 \rangle_{\rm PV}$ with hyperasymptotic approximation $(0,N_P)$, (4,0), $(4,31)$, (4,32), (4,33), (4,34) (black points). We also display the determination obtained in \cite{Bali:2014sja} (square blue point). For details see the main text.}
\label{Fig:LambdaFinal}
\end{figure}

{\bf *) Precision $(4,N')$}\\
We may try to increase the accuracy reached with the (4,0) hyperasymptotic approximation by adding the last term of \eq{eq:SPV}. Nevertheless, the errors quickly grow and get out of hand. This is mainly due to the error of the coefficients of the perturbative expansion. We have repeated the same error analysis than in the previous item for $N'=31$, 32, 33, 34. We show the obtained central values and errors in Fig. \ref{Fig:LambdaFinal}. We see how the errors quickly grow. Actually, the most important source to the error comes from the infinite volume extrapolation of the perturbative coefficients $p_n$.

\section{Discussion and conclusions}

In this paper, we have given the hyperasymptotic expansion of the plaquette with a precision that includes the terminant associated to the leading renormalon. Subleading effects are also considered. The perturbative series is regulated using the  PV prescription for its Borel integral. We use this analysis to give a determination of the gluon condensate in SU(3) pure gluodynamics: 
$$
\langle G^2 \rangle_{\rm PV} (n_f=0)=3.15(18)\, r_0^{-4}.
$$ 
We emphasize that this result is independent of the scale and renormalization scheme used for the coupling constant. Even if the computation was made in the lattice scheme, the result is the same in the $\MS$ scheme within the accuracy of the computation. 

$\langle G^2 \rangle_{\rm PV} (n_f=0)$ was computed with superasymptotic approximation in \cite{Bali:2014sja}. Here we have improved over this determination, principally by including the terminant associated to the leading renormalon. Adding $\Omega_{G^2}$ elliminates the jumps one has when using the superasymptotic approximation. The result is now much more smooth and flatter, to the point that, within errors, we can not isolate  ${\cal O}(a^2 \Lambda^2)$ effects. We still observe some small bending, which could also be due to higher-order perturbation theory. Overall, we are able to shrink the error by around a factor 2. 

In the lattice scheme, the impact of adding $\Omega_{G^2}$ is small compared with the size of the NP gluon condensate regulated using the PV prescription, of order 10\%. In the $\MS$ scheme, the contribution of the terminant would be larger by a factor $\sqrt{\al_{\MS}/\al_{latt}}$, which could easily enlarge the contribution by a factor 2. Note though that these statements are dependent on the value of $c$ used to fix $N_P$.

As we have seen in this paper, at present, the limiting factor for improving the determination of the gluon condensate in pure gluodynamics is the error of perturbation theory. All systematic sources of error have its origin in the errors of perturbation theory (even what we call statistical errors of \eq{eq:Fit} are dominated by the statistical errors of the coefficients $p_n$). More precise values of these perturbative coefficients, and its knowledge to higher orders, would yield a more precise determination of the normalization of the renormalons, $Z_P$, and would allow working with hyperasymptotic accuracy  $(4,N')$. Nowadays, if we try to reach this accuracy, we find that the error of the coefficients are too large to get accurate results. The situation with active light quarks is in an early stage but starts to be promising. The coefficients of the perturbative coefficients have been computed at finite volume in \cite{DelDebbio:2018ftu} for QCD with two massless fermions.  More data at different volumes, and the infinite volume extrapolation of these coefficients, would then allow to give a determination of the gluon condensate in QCD with two massless fermions. 

\begin{figure}
\centering
\includegraphics[width=0.95\textwidth]{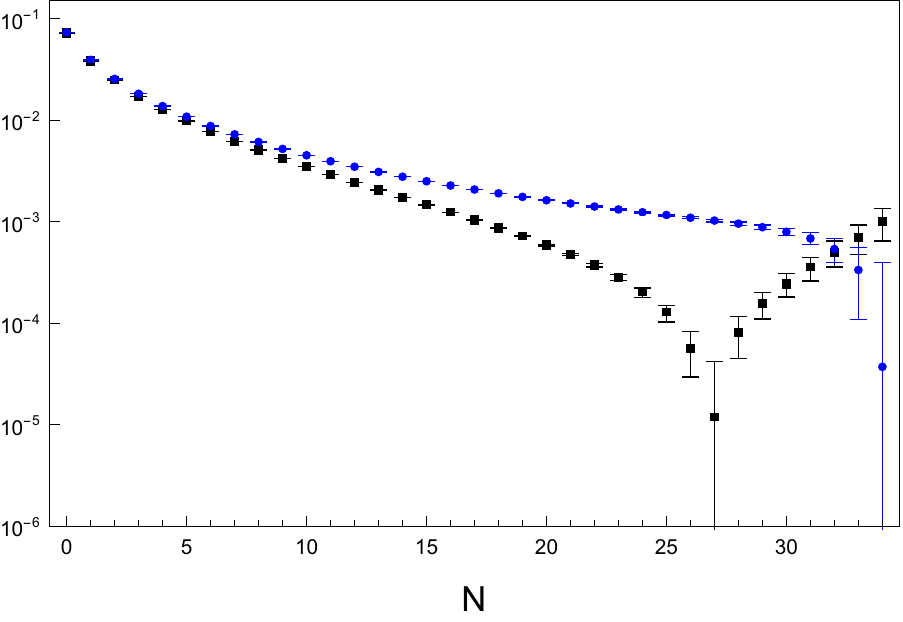}
\caption{We draw $\langle P \rangle_{\mathrm{MC}}-\sum_{n=0}^{N}p_n\al^{n+1}$ (blue points) and  $|\langle P \rangle_{\mathrm{MC}}-(\sum_{n=0}^{N}p_n\alpha^{n+1}+\frac{\pi^2}{36}C_{\rm G}(\al)\,a^4\langle G^2 \rangle_{\rm PV})|$ (black squares) for $\beta=6$ and $N \in [0,34]$. The error in all cases is the statistical error of the sum $\sum_{n=0}^{N}p_n\al^{n+1}$ and of $\langle P \rangle_{\rm MC}$ combined in quadrature. 
 Note that the plus/minus error does not display symmetrically in the plot because of the logarithmic scale, and also because of the logarithmic scale the error looks different for different points located at the same $N$.}
\label{Fig:6}
\end{figure}

\begin{figure}
\centering
\includegraphics[width=0.95\textwidth]{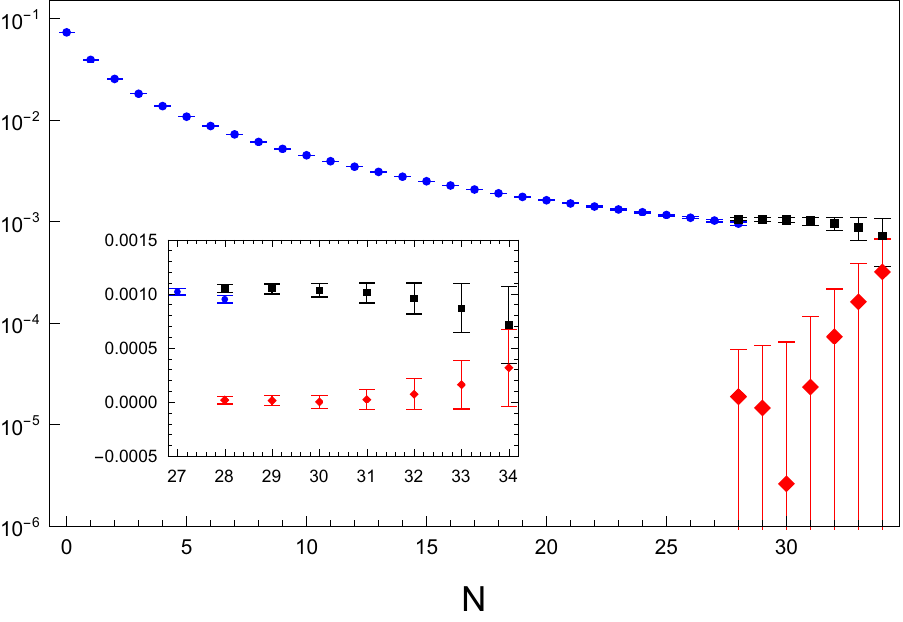}
\caption{We draw $\langle P \rangle_{\mathrm{MC}}-\sum_{n=0}^{N}p_n\al^{n+1}$ (blue points) for $\beta=6$ if $N\leq N_P$. For $ N= N_P$ we draw $\langle P \rangle_{\mathrm{MC}}-\sum_{n=0}^{N_P}p_n\al^{n+1}-\Omega_{G^2}$ (black squares). For $N> N_P$ we draw $\langle P \rangle_{\mathrm{MC}}-(\sum_{n=0}^{N_P}p_n\alpha^{n+1}+\Omega_{G^2}+
\sum_{n=N_P+1}^{N}[p_n-p_n^{\rm (as)}]\alpha^{n+1})$ (black squares). For $ N= N_P$ we draw $\langle P \rangle_{\mathrm{MC}}-(\sum_{n=0}^{N_P}p_n\al^{n+1}+\Omega_{G^2}+\frac{\pi^2}{36}C_{\rm G}(\al)\,a^4\langle G^2 \rangle_{\rm PV})$ (red diamonds). For $N> N_P$ we draw $\langle P \rangle_{\mathrm{MC}}-(\sum_{n=0}^{N_P}p_n\alpha^{n+1}+\Omega_{G^2}+\frac{\pi^2}{36}C_{\rm G}(\al)\,a^4\langle G^2 \rangle_{\rm PV}+
\sum_{n=N_P+1}^{N}[p_n-p_n^{\rm (as)}]\alpha^{n+1})$ (red diamonds). The error in all cases is the statistical error of the sum $\sum_{n=0}^{N}p_n\al^{n+1}$ and $\langle P \rangle_{\rm MC}$ combined in quadrature. 
 Note that the plus/minus error does not display symmetrically in the plot because of the logarithmic scale, and also because of the logarithmic scale the error looks different for different points located at the same $N$. In the small box a zoom of the points for $N \geq 27$ are shown in non-logarithmic scale.}
\label{Fig:hyp}
\end{figure}

It is interesting to show how our results fit general expectations for superasymptotic and hyperasymptotic expansions. Fig. \ref{Fig:6} nicely display, for a four-dimensional gauge theory, the standard behavior expected for perturbative series that are asymptotic to an observable (we take $\beta=6$ for illustrative purposes). We first discuss the blue points. First, as we add more terms to the perturbative series, such a perturbative series gets closer to the MC simulation of the plaquette. Nevertheless, as it is also expected for an asymptotic series, the rate of convergence diminishes till reaching an inflection point. This point is not a minimum. Adding extra terms to the perturbative series one approaches to the observable even if the perturbative series is divergent here. The reason one does not reach the minimum is the non-zero value of the gluon condensate using the PV prescription. If we also subtract the gluon condensate plus the perturbative expansion, we are in the same situation than in the large-$\beta_0$ models that were studied in \cite{Ayala:2019uaw,Ayala:2019hkn,Ayala:2019lak}, where the nonperturbative contribution is zero by construction. See Figs. 7 in \cite{Ayala:2019uaw,Ayala:2019hkn} for illustration. We then see the minimum (the maximal accuracy reached by the theory and how the series deteriorates if one continues adding extra perturbative terms). If at this points one adds $\Omega_{G^2}$ and the modified perturbative series where the leading renormalon is subtracted, one gets a plateau, and one can determine the gluon condensate. We illustrate this behavior for $\beta=6$ in Fig. \ref{Fig:hyp}. If one also subtracts the gluon condensate, as we do in this figure, one also has the jump in the precision achieved, as also observed in the large-$\beta_0$ plots (see again Figs. 7 in \cite{Ayala:2019uaw,Ayala:2019hkn} for illustration). 

We finally mention that the nonzero value of $\langle G^2 \rangle_{\rm PV}$ shows that the PV regularization of the perturbative sum, even if computed exactly, would differ from the Montecarlo simulation of the plaquette by a term of ${\cal O}(a^4\Lambda^4)$. This may affect the conjecture that the resummation technique of the perturbative expansion proposed in \cite{Caprini:2020lff} for the Adler function would not need such nonperturbative corrections. This should be further investigated.  

\medskip
 
{\bf Acknowledgments}\\
\noindent
This work was supported in part by the Spanish grants FPA2017-86989-P and SEV-2016-0588 from the ministerio de Ciencia, Innovaci\'on y Universidades, and the grant 2017SGR1069 from the Generalitat de Catalunya; and 
by FONDECYT (Chile) under grant No. 1200189. This project has received funding from the European Union's Horizon 2020 research and innovation programme under grant agreement No 824093. IFAE is partially funded by the CERCA program of the Generalitat de Catalunya.

\end{document}